\documentclass[10pt]{article}
\usepackage{bbm}
\usepackage{epsfig}

\usepackage{amsmath}
\usepackage{amsfonts,amsbsy}
\usepackage{amssymb}
\usepackage[dvips]{color}
\def\empile#1\over#2{\mathrel{\mathop{\kern 0pt#1}\limits_{#2}}}

\newcommand{\slv}{\raise.15ex\hbox{$/$}\kern-.53em\hbox{$v$}}
\newcommand{\slF}{\raise.15ex\hbox{$/$}\kern-.53em\hbox{$F$}}
\newcommand{\slL}{\raise.15ex\hbox{$/$}\kern-.53em\hbox{$L$}}
\newcommand{\slP}{\raise.15ex\hbox{$/$}\kern-.53em\hbox{$P$}}
\newcommand{\slp}{\raise.15ex\hbox{$/$}\kern-.53em\hbox{$p$}}
\newcommand{\slq}{\raise.15ex\hbox{$/$}\kern-.53em\hbox{$q$}}
\newcommand{\slR}{\raise.15ex\hbox{$/$}\kern-.53em\hbox{$R$}}
\newcommand{\slQ}{\raise.15ex\hbox{$/$}\kern-.53em\hbox{$Q$}}
\newcommand{\slK}{\raise.15ex\hbox{$/$}\kern-.53em\hbox{$K$}}
\newcommand{\slk}{\raise.15ex\hbox{$/$}\kern-.53em\hbox{$k$}}
\newcommand{\slD}{\raise.15ex\hbox{$/$}\kern-.53em\hbox{$D$}}
\newcommand{\slC}{\raise.15ex\hbox{$/$}\kern-.53em\hbox{$C$}}
\newcommand{\slA}{\raise.15ex\hbox{$/$}\kern-.53em\hbox{$A$}}
\newcommand{\slSigma}{\raise.15ex\hbox{$/$}\kern-.53em\hbox{$\Sigma$}}
\newcommand{\slpartial}{\raise.15ex\hbox{$/$}\kern-.53em\hbox{$\partial$}}
\newcommand{\slcalP}{\raise.15ex\hbox{$/$}\kern-.63em\hbox{$\cal P$}}

\def\p{{\boldsymbol p}}
\def\q{{\boldsymbol q}}

\def\k{{\boldsymbol k}}

\def\x{{\boldsymbol x}}
\def\y{{\boldsymbol y}}

\def\D{{\boldsymbol D}}

\def\bdel{{\boldsymbol \del}}
\def\balphat{\tilde {\boldsymbol \alpha}}

\def\alphat{{\tilde  \alpha}}

\def\betat{{\tilde  \beta}}

\def\At{{\tilde  A}}
\def\Ac{{\cal A}}

\newcommand{\beq}{\begin{eqnarray}}

\newcommand{\eeq}{\end{eqnarray}}

\long\def\comment#1{ }    

\newcommand{\be}{\begin{equation}}

\newcommand{\ee}{\end{equation}}

\newcommand{\nn}{\nonumber\\ }

\newcommand{\labe}{\label}

\def\del{\partial}

\catcode`\@=11

\newcount\@tempcntc
\def\@citex[#1]#2{\if@filesw\immediate\write\@auxout{\string\citation{#2}}\fi
  \@tempcnta\z@\@tempcntb\m@ne\def\@citea{}\@cite{%
        \@for\@citeb:=#2\do%
    {\@ifundefined{b@\@citeb}%
        {\@citeo\@tempcntb\m@ne\@citea%
                \def\@citea{,\penalty\@m\ }{\bf ?}\@warning%
                {Citation `\@citeb' on page \thepage \space undefined}}%
        {\setbox\z@\hbox{\global\@tempcntc0\csname b@\@citeb\endcsname\relax}
     \ifnum\@tempcntc=\z@ \@citeo\@tempcntb\m@ne%
       \@citea\def\@citea{,\penalty\@m}%
       \hbox{\csname b@\@citeb\endcsname}%
     \else%
      \advance\@tempcntb\@ne%
      \ifnum\@tempcntb=\@tempcntc%
      \else\advance\@tempcntb\m@ne\@citeo%
      \@tempcnta\@tempcntc\@tempcntb\@tempcntc\fi\fi}}\@citeo}{#1}}%

\def\@citeo{\ifnum\@tempcnta>\@tempcntb\else\@citea
  \def\@citea{,\penalty\@m}%
  \ifnum\@tempcnta=\@tempcntb\the\@tempcnta\else
   {\advance\@tempcnta\@ne\ifnum\@tempcnta=\@tempcntb \else
\def\@citea{--}\fi
    \advance\@tempcnta\m@ne\the\@tempcnta\@citea\the\@tempcntb}\fi\fi}

\catcode`\@=12

\begin{document}

\title{\bf The classical field created  in early stages of high energy nucleus-nucleus collisions }
\author{
Jean-Paul Blaizot $^1$,  Yacine Mehtar-Tani$^{2}$ }
\maketitle
\begin{center}
$^1$ ECT*, strada delle Tabarelle 286,
 I-38050 Villazzano (Trento), Italy\\
$^2$ Institut f\"ur Theoretische  Physik\\
Philosophenweg 16, D-69120 Heidelberg, Germany

\end{center}
\begin{abstract}

We show that a special choice of  light-cone gauge can greatly simplify the calculation of the classical  color field created in the initial stages of   nucleus-nucleus collisions. Within this gauge, we can in particular construct explicitly  the  conserved color current  and calculate exactly the gauge field immediately after the collision. This field is used as a boundary condition in an iterative solution of   the Yang-Mills equations in the forward light-cone. In leading order, which corresponds to a linearization of the Yang-Mills equation in the forward light-cone, we obtain a simple formula for the spectrum of gluons produced  in nucleus-nucleus collisions.   This  formula  reproduces exactly the known formula for proton-nucleus collisions, where $k_t-$factorization is recovered, while the latter property apparently breaks down in the case of nucleus-nucleus collisions.  

\end{abstract}

\vskip 5mm
\begin{flushright}
ECT*-08-06\\
HD-THEP-08-15
\end{flushright}

\section{Introduction}

It is believed that strong color fields are created in the initial stages of  nucleus-nucleus collisions at high energy.   In the color glass formalism, or its early formulation by  McLerran and Venugopalan \cite{McLerV} these fields are determined by solving the Yang-Mills equations for given distributions of color charges carried by the two nuclei. An average over these color charges is performed in the calculation of the observables. In the original formulation, the  distribution of the color charges carried by the   nuclei was assumed to be Gaussian. However, the definition of the color charges involves a separation of scales between the short wavelength partons, treated as sources, and the long wavelength ones, treated as fields. Independence of the final results upon this dividing scale leads to a renormalization group equation describing the non linear evolution of the parton density  in various forms  \cite{JalilKLW,KovneM1,KovneMW3,JalilKMW1,IancuLM,FerreILM1,Balit1,Kovch3}.

We shall not be concerned in this paper with this non linear evolution (for reviews see e.g. \cite{cargese,qgp3}), but shall focus merely on the determination of the initial classical field produced by a given distribution of color charges. This problem  has so far resisted a complete analytical treatment. Various numerical studies have been performed  \cite{KrasnV,KrasnNV,Lappi1}, based in particular on Refs.~\cite{JKMW1,JKMW2} (see also \cite{GV}), and several approaches have been suggested: Kovchegov's treatment relies on the conjecture that    final state interactions do not play any role, in addition to other assumptions \cite{KovAA}. More recently, Balitsky proposed a  symmetric expansion in powers of commutators of Wilson lines \cite{Balit2}. The only case for which a complete analytic treatment exists is that where one of the two sources can be considered as a weak color source. This is the situation for instance in proton-nucleus collisions. One can then  linearize  the Yang-Mills equations with respect the weak  source of the proton, while  treating all the high density effects in the strong source of the nucleus \cite{KovchM3,KovnW,KovchT1,DumitM1,BlaizGV1,GM}. This  allows one to express the cross-section for gluon production as a convolution in transverse momentum space of the  gluon distributions of the proton and the nucleus, a property known as  $k_t-$factorization.

 In this paper we report on progress in the analytic calculation of the color field produced in nucleus-nucleus collisions, and the resulting production of gluons.   We show that the complexity of the calculation is greatly reduced by an appropriate choice of gauge.  We choose an axial gauge that makes the field before the collision entirely concentrated in a narrow strip along  the light  cone. In particular, pure gauge fields that remain present in the forward light cone with most gauge choices, are absent  in the present formulation. This work generalizes to the case of nucleus-nucleus collision the study performed for the case of proton-nucleus collisions using a similar gauge choice  \cite{GM} .

    The paper is organized as follows. In Section 2, we study  the Yang-Mills equations in the light-cone gauge and show that important simplifications can be achieved by exploiting judiciously the remaining gauge freedom. We construct explicitly the  conserved current and calculate exactly the field immediately  after the collision.  Then we propose an iterative solution of the Yang-Mills equations in the forward light-cone. The leading order involves a linearization of the equations and is used in Section 3 to calculate the spectrum of gluons produced in nucleus-nucleus collisions. A  compact formula is obtained for this spectrum, which  does not exhibit $k_t$-factorization. The latter property is recovered in the case of proton-nucleus collisions  for which our formula yields the exact result. In Appendix \ref{FSG}  we show that our results can be recovered by using a variant of the  Fock-Schwinger gauge.


\section{Solving the Yang-Mills equations in light-cone gauge\label{LCGfield}}


We consider a collision of two nuclei,
denoted  A and B, that are moving at nearly the speed of light,
respectively in the $-z$ and the $+z$ directions, and carrying color
charge densities $\rho_{_A}$ and $\rho_{_B}$. Because of the Lorentz
contraction, these densities have a small longitudinal extent, which translates into   small extents, of order $\epsilon$, in the $x^+$ direction for the nucleus A and in the $x^-$ direction for the nucleus B, where $x^\pm=(t\pm z)/\sqrt{2}$  denote the light cone variables (see Fig.~\ref{fig1}). Thus $\rho_{_A}$ and $\rho_{_B}$ are functions of $x^+$ and $x^-$, respectively $\rho_{_A}(x^+,\x)$ and $\rho_{_B}(x^-,\x)$, with $\x$ the transverse coordinates. 

Note that  the integrals of $\rho_{_A}$ over $x^+$, and $\rho_{_B}$ over  $x^-$, are finite; it follows that the color charge densities diverge  as $1/\epsilon$ when $\epsilon\to 0$. One often takes  the limit $\epsilon\rightarrow 0$
from the outset, writing $\rho_{_{A,B}}\propto \delta(x^{\pm})$.
However, the longitudinal structure of the sources  plays a role in the average over the color sources \cite{Lappi:2007ku,Fukushima:2007ki} and also, as we shall see, in the analysis of the Yang-Mills equations, in particular in the determination of  the continuity conditions of the fields at the boundary of the nuclei. We shall  therefore    keep $\epsilon$ finite and take the high energy limit
($\epsilon\to 0$) only at the end of the calculations.

The classical gauge field is solution of the Yang-Mills equations 
\be\label{Yang-Mills}
D_\mu F^{\mu\nu}=J^{\nu},
\ee
where $D_\mu=\del_\mu-igA^c_\mu T^c\equiv\del_\mu-igA_\mu\cdot T$ is the covariant derivative, $T^c$ is a generator of $SU(3)$ in the adjoint representation, and  $A_\mu^a$, $F_{\mu\nu}^a$ and  $J_\mu^a$ are components of color vectors (of the adjoint representation). Equations such as Eq.~(\ref{Yang-Mills}) are to be read as  equations between color vectors (whose color index is omitted to alleviate the notation). The color current satisfies the conservation equation
\beq\label{DJ}
D_\mu J^\mu=0\,.
\eeq
In a gauge transformation, 
\beq
A^\mu\cdot T\to \Omega(A^\mu \cdot T)\Omega^\dagger-\frac{1}{ig}\Omega\partial^\mu \Omega^\dagger,
\eeq
with $\Omega$ a  matrix in the adjoint representation of SU(3), the color current transforms as
\beq
J^\mu \to \Omega J^\mu.
\eeq
Before the collision, the only components of the color current are
\beq
J^+(x)=\rho_{_B}(x^-,\x),\qquad J^-(x)=\rho_{_A}(x^+,\x).
\eeq
Note that the precise definition of the charge densities, and the color currents, involves a choice of gauge: we shall assume throughout that the color densities are those of the covariant gauge (see Sect.~2.1).

In most of this paper, we shall work in the light-cone gauge $A^+=0$. The
Yang-Mills equations read then \cite{GM} \beq &&
-\partial^+(\partial_\mu A^\mu)-ig(A^i\cdot T)\;\partial^+A^i=J^+, \labe{YM1}
\\
&& D^-\partial^+A^--D^iF^{i-}=J^-,\labe{YM2}
\\
&&
\partial^+F^{-i}+D^-\partial^+A^i-D^jF^{ji}=J^i.\labe{YM3}
\eeq Note that the first equation, Eq.~(\ref{YM1}),  does not contain any
 derivative with respect to $x^+$ ($\partial^-=\partial /\partial x^+$).
Therefore, it can be seen as a constraint that relates the various
field components at the same $x^+$. 

\subsection{The gauge field before the collision\label{initialcond}}


Before the collision, i.e., for $t<0$, the gauge field 
is obtained by solving the Yang-Mills equations for each of the
projectiles separately. 
We shall denote by $A^\mu_{_A}$ ($A^\mu_{_B}$) the field created by the nucleus A (B) in the absence of the nucleus B (A).

Consider first the nucleus A. The gauge field that it produces is not
completely fixed by the gauge condition $A^+=0$. One can exploit the residual gauge freedom to choose  the gauge field
at $x^-<0$   to be
 either    longitudinal   or   transverse.

The longitudinal solution is obtained from Eq.~(\ref{YM2}), by assuming  that the component $A^-$ does not depend on $x^-$, i.e., $\partial^+A^-=0$, and that the transverse components vanish, $A^i=0$ \cite{GM}. One gets then 
\be\labe{Afield-}
A^-_{_A}(x)=\Phi_{_A}(x^+,\x),\qquad A^i_{_A}=A^+_{_A}=0 , 
\ee 
where 
\be\label{Phi_A}
-\bdel_\perp^2\Phi_{_A}(x^+,\x)=\rho_{_A}(x^+,\x).
\ee

With this gauge choice, the field created by the nucleus A before the collision has support  in a small strip of width $\epsilon$ along the semi-axis $x^{-}<0$ (see Fig.~\ref{fig1} below). The corresponding  current created by the nucleus A is simply
\beq\label{JmuA}
J^\mu_{_A}(x)=\delta^{\mu -}\rho_{_A}(x^+,\x).\labe{j-in}
\eeq
It is conserved since 
\be\label{conservedJA}
D_\mu J^\mu_{_A}=\left(   \partial^+-ig A^+_{_A}\cdot T\right)  J^-_{_A}=\partial^+J^-_{_A}=0,
\ee
where we have used successively the gauge condition $A^+_{_A}=0$ and the fact that $J^-_{_A}$ is independent of $x^-$.

The transverse   solution can be obtained by performing a gauge transformation on (\ref{Afield-}) that eliminates $A^-$. Denoting by $U^\dagger$ the matrix that realizes the gauge transformation, we have
\beq
&& A^{ +}_{_A}\cdot T \to -\frac{1}{ig} U^\dagger \partial ^+ U=0,\nonumber\\
&& A^-_{_A}\cdot T \to U^\dagger (\Phi_{_A}\cdot T)\;U-\frac{1}{ig} U^\dagger\partial ^- U=0,\nonumber\\
&& A^{i}_{_A}\cdot T\to -\frac{1}{ig} U^\dagger\partial ^i U.\label{gaugerotation}
\eeq
We shall verify shortly that the solution of these equations is provided by the following  Wilson line in the adjoint representation:
\be\label{WilsonU1} 
U(x^+,\x)\equiv {\mathcal
P}\exp{\left[ig\int_{-\infty}^{x^+}dz^+\Phi_{_A}(z^+,\x)\cdot T\right]}.
\ee 
The symbol ${\mathcal P}$ in Eq.~(\ref{WilsonU1}), and throughout, denotes path ordering (along the integration variable, here $z^+$). Note that the
integral in the  Wilson line  (\ref{WilsonU1}) gets contribution only within the
support of $\rho_{_A}$, i.e., for $0<z^+<\epsilon$. Therefore, $U$ becomes independent of $x^+$ when $x^+>\epsilon$, and $U(x^+<0)=1$.

With $U$ given by Eq.~(\ref{WilsonU1}), the second equation  (\ref{gaugerotation}) yields $A^-=0$. The first equation (\ref{gaugerotation}) ensures that the gauge condition $A^+=0$ is satisfied since $U$ does not depend on $x^-$. The only non vanishing components of $A$ are, as announced, the transverse components $A^i $, which we may rewrite as follows
\be
A^i_{_A}(x)=-\int_{-\infty}^{x^+}dy^+U^\dag(y^+,\x)\;\del^i\Phi_{_A}(y^+,\x), \labe{Afieldvec}
\ee
where we have used   the following identity (valid for any matrix $U$ of the adjoint representation):
\beq\label{identity}
U^\dagger T^a U=U_{ab}T^b= T^b U^\dagger_{ba}.
\eeq
Note that since $U=1$ for $x^+<0$, the field $ A^i_{_A} $ is non zero only in the region $x^+>0$. Note also that $A^i_{_A}$ is independent of $x^+$ when $x^+>\epsilon$. It follows that  $A^i_{_A}$ becomes, in the limit $\epsilon\to 0$,  a discontinuous function of $x^+$, at $x^+=0$. 

The color current corresponding to this transverse solution is related to the current (\ref{JmuA}) by the same gauge transformation as in Eq.~(\ref{gaugerotation}):
\beq\label{JmuA2}
J^\mu_{_A}(x)\to U^\dag  J^\mu_{_A}(x)=\delta^{\mu-}U^\dag(x^+,\x) \rho_{_A}(x^+,\x).\labe{JA}
\eeq
This equation makes explicit the dependence of the color density, and the associated color current, on the choice of the gauge.  Note that  the longitudinal
 solution (\ref{Afield-})  is common to both the light-cone gauge  $A^+=0$ and the covariant gauge \cite{BlaizGV1}: indeed, when $x^-<0$,  $\del_\mu A_{_A}^\mu=\del^+ A^-_{_A}=0$ when $A_{_A}^\mu$ satisfies (\ref{Afield-}). This is the reason for the simple form (\ref{JmuA}) of the current for the longitudinal solution. In most of this paper, we shall use  the longitudinal solution (\ref{Afield-}) as the field produced by the nucleus A. 

Let us turn now to the nucleus   B. A {\it priori} there exists for B the same gauge freedom as for  A. However the condition $A^+=0$ forces us here to choose the transverse solution (see however Sect.~2.5). With $\rho_{_B}(x^-,\x)$ the color charge of the nucleus B (in covariant gauge), and $\Phi_{_B}(x^-,\x)$ the solution of (cf. Eq.~(\ref{Phi_A}))
\be\label{Phi_B}
-\bdel_\perp^2\Phi_{_B}(x^-,\x)=\rho_{_B}(x^-,\x),
\ee
we obtain  
\be
A^{i}_{_B}\cdot T=-\frac{1}{ig}V^\dag(x^-,\x)\del^iV(x^-,\x)\;\;\text{and}\;\;A_{_B}^+=A_{_B}^-=0, \label{Bfieldt}
\ee
where $V$ is a Wilson line along $x^-$:
\be\label{WilsonV1} 
V(x^-,\x)\equiv {\mathcal
P}\exp{\left[ig\int_{-\infty}^{x^-}dz^-\Phi_{_B}(z^-,\x)\cdot T\right]}.
\ee
More explicitly,  
\beq\label{AiB}
A^{i}_{_B}(x)=-\int_{-\infty}^{x^-}dy^-V^\dag(y^-,\x)\;\del^i\Phi_{_B}(y^-,\x). \labe{Bfieldvec}
\eeq
Note that since $\Phi_{_B}(x^-)$ has support in the strip $0<x^-<\epsilon$, $A^{i}_{_B}(x)$ vanishes for $x^-<0$ and is independent of $x^-$ for $x^->\epsilon$: as $\epsilon\to 0$ a discontinuity builds up in $A^{i}_{_B}(x^-)$

The  current of the nucleus B is related to $\rho_{_B}$ by (cf. Eq.~(\ref{JmuA2}))
\be
J^{\mu}_{_B}(x)=\delta^{\mu+}V^\dag(x^-,\x) \rho_{_B}(x^-,\x).\labe{j+in}
\ee
This current is conserved since 
\be
D_\mu J^\mu_{_B}=
\left( \partial^--ig A_{_B}^-\cdot T\right)J^+_{_B}=\partial^-J^+_{_B}=0,
\ee
where we have used successively $A^-_{_B}=0$ and the fact that $J^+_{_B}(x)$ is independent of $x^+$.

\begin{figure}[ht]
\begin{center}
\resizebox*{!}{6cm}{\includegraphics{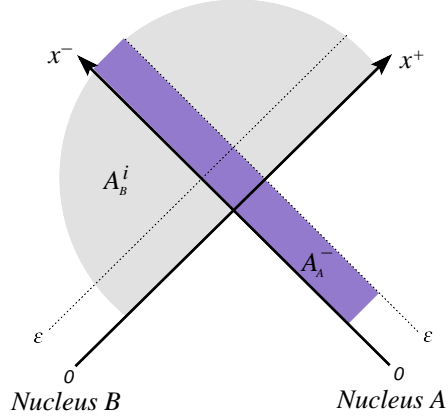}}
\caption{The gauge field before the collision, in the light-cone gauge $A^+=0$. The nucleus A produces a static longitudinal field localized in a strip of width $\epsilon$  near the light cone along the  $x^-$ axis. The field of the nucleus B   is transverse and exists in the half plane $x^->0$. It is independent of $x^\pm$ for $x^->\epsilon$. }\label{fig1}
\end{center}
\end{figure}


\subsection{The conserved current \label{current}}


We have verified in the previous subsection that the currents $J^\mu_{_A}$ and $J^\mu_{_B}$,  carried respectively by the nuclei A and B, are separately conserved before the interaction. In this subsection we shall obtain the expression of the conserved current for the interacting system. To this aim we note first that the conservation law $D_\mu J^\mu=0$, Eq.~ (\ref{DJ}),   constrains  only one component of the current. We shall exploit the remaining freedom to force the current to satisfy the extra relations:
\beq\label{DJ2}
D^+ J^-=0, \qquad D^- J^+=0, \qquad J^i=0.
\eeq
A current satisfying (\ref{DJ2}) obviously satisfies (\ref{DJ}). Note also that the relations (\ref{DJ2}) are compatible with the conservation laws  (\ref{j-in}) and (\ref{j+in}) satisfied by the current before the interaction. 
Thanks to the gauge condition $A^+=0$, the first equation in (\ref{DJ2}) reduces to
\be\label{J-}
\partial^+J^-=0.
\ee
This implies that the current $J^-$ is conserved along the $x^-$ direction, and is therefore given by the initial current at $x^-\rightarrow -\infty$, namely  
\beq\label{conservedJA2}
J^-(x)=J^-_{_A}(x^+,\x),
\eeq
with $J^-_{_A}$ the current of the nucleus A alone, given in Eq.~(\ref{j-in}).
The second equation, $D^-J^+=0$, is solved by 
\be\label{J+}
J^+(x)=W(x)\;J^+_{_B}(x^-,\x),
\ee
with 
\be\label{WlineW}
W(x)\equiv {\mathcal
P}\exp{\left[ig\int_{-\infty}^{x^+}dz^+A^-(z^+,x^-,\x)\cdot T\right]}.
\ee
The Wilson line involves  the component $A^-(x)$ of the full solution of the Yang-Mills equations, which is {\it a priori} not known. However, the  current $J^+_{_B}$ in Eq.~(\ref{J+}), proportional to $\rho_{_B}(x^-,\x)$, has support only in the narrow strip  $0<x^-<\epsilon$ along the $x^+$ axis. Therefore, in  the limit $\epsilon\to 0$, only the value of $A^-$ along the $x^+$ axis is needed to evaluate $W$.  
In Appendix \ref{Proof1} we show that this   is given by the field $\Phi_{_A}$ created by  the nucleus A alone:
\be
\lim_{\epsilon\rightarrow 0}A^-(x^+,x^-=\epsilon,\x)\equiv \Phi_{_A}(x^+,\x).\labe{A-0}
\ee
Thus, in the limit $\epsilon\to 0$, the Wilson line $W$ of Eq.~(\ref{WlineW}) reduces  to
the Wilson line $U(x^+,\x)$ given in Eq.~(\ref{WilsonU1}).

It follows that, in the light-cone gauge $A^+=0$,  the conserved current in a nucleus-nucleus collision can be written thus 
\beq\label{Jmu}
J^\mu(x)=\delta^{\mu-}\rho_{_A}(x^+,\x)+\delta^{\mu+}U(x^+,\x) V^\dag(x^-,\x) \rho_{_B}(x^-,\x),
\eeq
The first Wilson line acting on $\rho_B$ brings the density from the covariant gauge to the light-cone gauge $A^+=0$. The second Wislon line, $U(x^+,\x)$ summarizes the effect of the interaction.
The formula  (\ref{Jmu}) shows that, in the light-cone gauge, the conserved current can be constructed explicitly in terms of the  sources $\rho_{_A}$ and $\rho_{_B}$ (defined in covariant gauge). In the interaction zone a color rotation takes place, but this is determined by the gauge fields created by the individual nuclei alone, and these are known analytically. In contrast, in covariant gauge, the determination of the conserved current requires the solution of the Yang-Mills equation in the interaction zone and this can be done only approximately \cite{BlaizGV1}. Note that the interaction leaves  the component $J^-$ unaffected (as a consequence of (\ref{conservedJA2})), while  $J^+$ acquires an $x^+$ dependence through the gauge rotation. There is therefore an apparent dissymmetry in the present description that has, however, no consequence in the calculation of the gluon production, as we shall verify explicitly in Sect.~\ref{gluprod}.
Had we chosen for the field of the nucleus A the  transverse solution (\ref{Afieldvec}), one would have obtained a more symmetrical current  $J^\mu\rightarrow U^\dag J^\mu=\delta^{\mu-}U^\dag\rho_{_A}+\delta^{\mu+} V^\dag \rho_{_B}$, similar to that obtained  in the Fock-Schwinger gauge $x^+A^-+x^-A^+=0$; this leads to the same spectrum of produced gluons (see Appendix \ref{FSG} for a more complete discussion of the Fock-Schwinger gauge).


\subsection{The gauge field immediately after the collision\label{fieldborder}}

\begin{figure}[ht]
\begin{center}
\resizebox*{!}{6cm}{\includegraphics{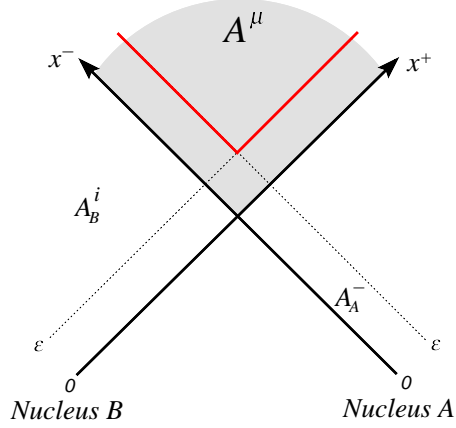}}
\caption{ The field $A^\mu$ in the forward light-cone is obtained by solving the Yang-Mills equations with initial conditions given on the border of a strip of width $\epsilon$ along the light-cone (indicated by the thick (red) line). The field in this strip is determined analytically as a function of the source $\rho_{_A}$ and $\rho_{_B}$ in subsection 2.3. }\label{fig2}
\end{center}
\end{figure}

We now turn to the determination of the gauge field  immediately after the collision, that is,  along the lines $x^+= \epsilon$ or $x^-= \epsilon$ (indicated by a thick (red) line in Fig.~\ref{fig2}), in the limit where $\epsilon\to 0$.  As we shall see, this field  can be determined analytically.  It will serve as a boundary value  for the solution of the Yang-Mills equations  in the forward light-cone  $x^\pm>\epsilon$, that will be considered in the next subsections.

Before we start our analysis, it is useful to recall that the densities $\rho_{_A}(\x^+,\x)$ and $\rho_{_B}(x^-,\x)$ diverge as $1/\epsilon$ when $x^\pm=\epsilon\to 0$. The same holds for the fields $\Phi_{_A}(\x^+,\x)$ and $\Phi_{_B}(x^-,\x)$, solutions of Eqs.~(\ref{Phi_A}) and (\ref{Phi_B}), respectively. The transverse components of the field remain finite, however, as can be seen for instance on Eq.~(\ref{Afieldvec}) for $A^i_{_A}$: the divergence $\sim 1/\epsilon$ of $\Phi_{_A}$ is eliminated by the $y^+$ integration $\sim \epsilon$. We shall now proceed to an analysis of the Yang-Mills equations in the region $x^\pm\le \epsilon$, keeping only the dominant terms as $\epsilon\to 0$.

By integrating the  Yang-Mills equation (\ref{YM3}) over $x^-$ from 0 to $\epsilon$, at fixed $x^+$ and $\x$, we get
\beq
&& 2\del^-A^i(\epsilon)-\del^i[A^-(\epsilon)-\Phi_{_A}]-ig(A^-(\epsilon)\cdot T)\;A^i(\epsilon)
\nn
&&\quad\quad-ig\int_{0}^\epsilon dx^- (A^-\cdot T)\;\del^+A^i(x^-) -\int_{0}^\epsilon dx^- D^j F^{ji}(x^-)=0.\nn\labe{alphai}
\eeq
Simplifications occur in the limit $\epsilon\to 0$: 
the second term  drops out because of Eq.~(\ref{A-0});  the last  term  vanishes  since $F_{ij}$ and its derivatives with respect to the transverse coordinates remain finite as $\epsilon\to 0$; finally, using again the continuity of $A^-(x^-)$, Eq.~(\ref{A-0}), one can perform the integral in the fourth term and show that its contribution is identical to that of the third one. Thus, Eq.~(\ref{alphai}) reduces to 
\be
\del^-A^i(x^+,\epsilon,\x)-ig\,[\Phi_{_A}(x^+,\x)\cdot T]\,A^i(x^+,\epsilon,\x)=0.
\ee
This equation governs the $x^+$ dependence of the transverse components of the field, along the axis $x^-=\epsilon$. It is valid in fact for all $x^-\le \epsilon$. It is easily solved, to yield: 
\beq\label{Aidex-}
A^i(x)=U(x^+,\x)A^i_{_B}(x^-,\x),\qquad x^-\le \epsilon,
\eeq
with $U$ given by Eq.~(\ref{WilsonU1}) and $A^i_{_B}$ by Eq.~(\ref{Bfieldvec}). We have used the fact that  $A^i(x^+=0,x^-,\x)=A^i_{_B}(x^-,\x)$. 

Similarly, by integrating Eq.~(\ref{YM1}) over $x^-$ up to $\epsilon$, at fixed $x^+$ and $\x$, we obtain
\be
-\del^+A^-(\epsilon)+\del^iA^i(\epsilon)-ig\int_0^\epsilon dx^-(A^i\cdot T)\;\del^+ A^i(x^-)=\int_0^\epsilon dx^-UV^\dag\rho_{_B}(x^-).
\ee
Replacing in this equation the field $A^i$ by its explicit expression (\ref{Aidex-}) and performing simple manipulations (using in particular Eq.~(\ref{identity})), one can rewrite this equation as
\beq
 &&\del^+A^-(\epsilon)=(\del^iU)A_{_B}^i(\epsilon)\nn
&&\quad+U\left(\del^iA_{_B}^i(\epsilon)-\int_0^\epsilon dx^- (A_{_B}^i\cdot T)\;\del^+A_{_B}^i(x^-)-\int_0^\epsilon dx^-V^\dag\rho_{_B}(x^-)\right).\labe{beta02}\nn
\eeq
The terms between parentheses  that multiply $U$ in the above equation, vanish since $A_{_B}^i$ solves the Yang-Mills equation  (\ref{YM1}) in the absence of the nucleus A. Therefore, we are left with
\be
\del^+A^-(x)=\left(\del^iU(x^+,\x)\right)A_{_B}^i(x^-,\x),\qquad x^-\le \epsilon  .\labe{A-dex-}
\ee

We show now that the exact solutions (\ref{Aidex-}) and (\ref{A-dex-}) that are valid in the strip $0\le x^-\le\epsilon$, extend to the region  $0\le x^+\le\epsilon$. To do that, we consider Eq.~(\ref{YM3}) again, but now for $x^+\le\epsilon$, and recall that, in the limit  $\epsilon\to 0$, $\Phi_{_A}(x^+,\x)\sim 1/\epsilon$. By keeping the dominant terms in the limit of small $\epsilon$,  one can then reduce Eq.~(\ref{YM3}) to the following equation 
\beq
\del^- \del^+ A^i(x)-ig\,[\Phi_{_A}(x^+,\x)\cdot T]\,\del^+A^i(x)=0,
\eeq
valid for all $x^+\le\epsilon$. This equation is easily solved and yields
\beq
\del^+ A^i(x)=U(x^+,\x) \del^+ A^i_{_B}(x^-,\x),
\eeq
where we have used the fact that $A^i(x^+=0,x^-,\x)=A^i_{_B}(x^-,\x)$. A last integration over $x^-$, using $A^i(x^-=0)=A^i_{_B}( x^-=0)=0$ (see Eqs.~(\ref{Aidex-}) and (\ref{AiB})),  gives finally
\beq\label{Aidex+}
A^i(x)= U(x^+,\x) A^i_{_B}(x^-,\x),\qquad x^+\le \epsilon,
\eeq
which is identical to $(\ref{Aidex-})$, but now valid  also in the strip $x^+\le \epsilon$.

Analogous manipulations allow us to transform Eq.~(\ref{YM2}) into the following equation for $\del^+A^-(x)$, valid for $x^+\le \epsilon$:
\beq\label{del-del+a-}
\left(\del^-  -ig \;\Phi_{_A}\cdot T\right)  \del^+A^-=-ig\;(A^i\cdot T )\; \del^i\Phi_A,
\eeq
where we have used Eq.~(\ref{conservedJA2}),   (\ref{JmuA}) and (\ref{Phi_A}) in order to eliminate the current $J^-$.  At this point we use the identity
\beq
\del^-\left[ U^\dagger(x^+,\x)\; \del^+A^-(x) \right] =U^\dagger(x^+,\x) \left[ \del^--ig\;\Phi_{_A}(x^+,\x)  \right] \del^+A^-( x),
\eeq
and rewrite Eq.~(\ref{del-del+a-}) as
\beq
\del^-\left[ U^\dagger(x^+\x)\;\del^+A^-(x) \right]=-ig U^\dagger(x^+,\x)\;(A^i (x)\cdot T)\; \del^i\Phi_{_A}(x^+,\x).
\eeq
By integrating over $x^+ \le \epsilon$, using $A^-( x^+=0)=0$, one gets
\beq\label{A-dex+}
 \del^+A^-( x)&=&-ig U(x^+,\x)\int_0^{x^+} dz^+ U^\dagger(z^+,\x)\;(A^i(z^+)\cdot T)\;\del^i\Phi_{_A}(z^+,\x)\nn
&=& igU(x^+,\x)\int_0^{x^+} dz^+ U^\dagger(z^+,\x) \del^i(\Phi_{_A}(z^+)\cdot T)A^i(z^+)\nn
&=&(\del^iU(x^+))A_{_B}^i(x^-),
\eeq
where in the last step we have used the expression (\ref{Aidex+}) of $A^i(x)$. 
Eq.~(\ref{A-dex+}) is identical to Eq. (\ref{A-dex-}), but now valid in the strip $0\le x^+\le \epsilon$.
 
Summarizing, we have shown that the field components in the strips   $0\le x^+\leq \epsilon $ or $0\le x^-\leq \epsilon$, are given by 
\be\labe{Astrip}
A^i(x)=U(x^+,\x)A^i_{_B}(x^-,\x),\qquad A^-(x)=\Phi_{_A}(x^+,\x),\qquad A^+=0,
\ee
and 
\be
\del^+A^-(x)=\left(\del^iU(x^+,\x)\right) A_{_B}^i(x^-,\x),
\ee
where $A^i_B(x^-,\x)$ is given by Eq.~(\ref{Bfieldt}). 
Since $V(x^-\geq \epsilon,\x)=V(x^-=\epsilon,\x)\equiv V(\x)$ , $A^i_{_B}(x^-\geq\epsilon,\x)\equiv A^i_{_B}(\x)$ is constant in the forward light-cone $x^\pm>\epsilon$. The same holds for $U$: $U(x^+\geq \epsilon,\x)=U(x^+=\epsilon,\x)\equiv U(\x)$. It follows that  at the border of  the small strip of width $\epsilon$ along the light-cone, the field is constant (independent of $x^+$ and $x^-$) and given by 
\beq\labe{Aborder}
&& A^i(\x)=U(\x)A^i_{_B}(\x),\qquad A^+=A^-=0\;,\nn
&&\del^+A^-(\x)=\left(\del^iU(\x)\right) A_{_B}^i(\x).
\eeq

This solution is related to the choice of the longitudinal field (\ref{Afield-}) for the nucleus A. We have seen that by performing the gauge rotaton (\ref{gaugerotation}), which does not affect the light-cone gauge condition $A^+=0$, one obtains a   transverse field for the nucleus A, which is analogous to the field of the nucleus B (cf. Eq. (\ref{Afieldvec}) and Eq. (\ref{Bfieldvec}), respectively). Since, in this case $A^+=A^-=0$,  the field before the collision is then similar to that in the Fock-Schwinger gauge $x^+A^-+x^-A^+=0$. 
To make the comparison more explicit, we note that the field produced on the light-cone by the initial transverse field of Eqs. (\ref{Afieldvec}) and (\ref{Bfieldvec}), is obtained from (\ref{Astrip}) by performing the gauge rotation (\ref{gaugerotation}) that  eliminates $A^-=\Phi_{_A}$. This  yields (at the border of the strip where the field  becomes  independent of $x^+$ and $x^-$):
\beq\labe{Aborder2}
&&A^i(\x)=A^i_{_B}(\x)+A^i_{_A}(\x)\;,\;  A^+=A^-=0,\nn
&&\del^+A^-(\x)=(A^i_{_A}(\x)\cdot T) A_{_B}^i(\x).
\eeq  
These field components on the light-cone may be compared to the corresponding ones obtained in the Fock-Schwinger gauge (see Eqs. (22) and (23) of Ref. \cite{JKMW1}). The transverse  components, first line of Eq. (\ref{Aborder2}), are identical to those in the Fock-Schwinger gauge, while the derivative  $\del^+A^-$ is related to the transverse field by a similar equation, second line of Eq. (\ref{Aborder2}), to within a   factor $1/2$ (see also Eq. (\ref{FSborder}) in Appendix \ref{FSG} where a more detailed analysis is presented). 

In the next subsections we shall consider the field in the forward light-cone, and present a method for solving the Yang Mills equations.  We shall see that having  transverse, pure gauge fields, for the nuclei A and B, whose supports extend respectively to the half planes $x^+>0$ and $x^->0$, leads to spurious ``final state interactions'' in the forward light-cone, all the way till $t=+\infty$. The presence of these pure gauge components greatly complicates the calculation of observables, and it is desirable to get rid of them. We shall see that this is possible by a slight redefinition of the gauge condition, which will be discussed in the subsection \ref{newgauge}.

\subsection{The gauge field in the forward light-cone \label{forwardLC}}

While it has been possible to calculate exactly the gauge field on the light-cone, the corresponging calculation  
  in the forward light-cone remains a hard problem, because of the non linearity of the Yang-Mills equations. Our strategy will then be to construct the solution through an iterative procedure. This produces a solution in the form of an expansion: 
\be
A^\mu=\sum_{n=0}^{\infty}A^\mu_{(n)},\labe{expan}
\ee
where $A^\mu_{(0)}$ corresponds to the non-interacting part of the gauge field, i.e,
\be\label{A0}
A^i_{(0)}=A^i_{_B}, \qquad A^-_{(0)}=\Phi_{_A}, \qquad A^+_{(0)}=0.
\ee
The first correction $A^\mu_{(1)}$ is obtained by solving the linearized Yang-Mills equations  (in $A^\mu_{(1)}$, keeping $A^\mu_{(0)}$ as a background field) in the forward light-cone, using as initial conditions the field on the light-cone, which is known exactly. The process is repeated for $A^\mu_{(2)}$, and so on. Note that the boundary condition on the border of the strip  $\{x^+=\epsilon, x^->\epsilon ;x^-=\epsilon,x^+>\epsilon\}$ reads $A^\mu_{(n)}=0$ for $n>1$. That is, near the light cone, the field is entirely given by 
$A^\mu_{(0)}+A^\mu_{(1)}$.  There is therefore some analogy between the expansion (\ref{expan}) and the proper time expansion introduced in Ref.~\cite{Fries}, but the two expansions are distinct.

 In this subsection we shall construct $A^\mu_{(1)}$ explicitly, and show that its calculation can be greatly simplified by a convenient choice of gauge. We shall introduce the following notation for the   fields on the light-cone (Eqs.~(\ref{Aborder})):
\beq
&&\alpha^i_{_0}(\x)\equiv A^i(\x)-A_{_B}^i(\x)=(U(\x)-1)A^i_{_B}(\x),\label{alpha0}\\
&& \beta_{_0}(\x)\equiv \del^+A^-(\x)=\left(\del^iU(\x)\right) A_{_B}^i(\x).\label{beta0}
\eeq

Let us consider then the Yang-Mills equations in the forward light-cone, i.e., for  $x^\pm >\epsilon$. The linearized  equations read 
\be
D_{(0)\mu} F_{(1)}^{\mu\nu}+D_{(1)\mu} F_{(0)}^{\mu\nu}=0,
\ee
where $D_{(0)}^\mu=\del^\mu-ig A_{(0)}^\mu\cdot T$, and  $A_{(0)}^\mu$ is given in Eq.~(\ref{A0}) above. 
Since $\Phi_{_A}$ is confined to the strip $0<x^+<\epsilon$, only $A^i_{_B}$ contributes to $A^\mu_{(0)}$ and therefore  $D_{(0)}^+=\del^+$, $D_{(0)}^-=\del^-$ and $D_{(0)}^i=\del^i-ig A_{_B}^i\cdot T$. Furthermore, since $A^i_{_B}$ is a pure gauge,  $F_{(0)}^{\mu\nu}=0$,  and the second term vanishes. We are left with
\be
D_{(0)\mu} D_{(0)}^\mu A_{(1)}^\nu- D_{(0)}^\nu D_{(0)\mu} A_{(1)}^\mu=0.
\ee 
Making the components explicit, we get
\beq
&&\!\!\!\!\del^+\left[ \del^+A^-_{(1)}-D_{(0)}^i A_{(1)}^i\right]=0,\labe{YMlin+}\\
&&\!\!\!\! \del^+\del^-A^-_{(1)} -D_{(0)}^iD_{(0)}^iA^-_{(1)}+\del^-D_{(0)}^iA^i_{(1)}=0,\labe{YMlin-}\\
&&\!\!\!\!2\del^+\del^-A^i_{(1)} -D_{(0)}^jD_{(0)}^jA^i_{(1)}+D_{(0)}^iD_{(0)}^jA^j_{(1)}-D_{(0)}^i\del^+A^-_{(1)}=0.\labe{YMlint}
\eeq
By integrating  Eq. (\ref{YMlin+}) over $x^-$,  from $\epsilon$ to $x^->\epsilon$ we obtain 
\be
\del^+A^-_{(1)}-D_{(0)}^iA^i_{(1)}=\beta_{_0}-D_{(0)}^i\alpha^i_{_0},\labe{const2}
\ee
where we have used the fact that $\del^+A^-(x^-=\epsilon)=\beta_{_0}$ and $A^i(x^-=\epsilon)=\alpha^i_{_0}$. By using this relation  (\ref{const2}) in Eq.~(\ref{YMlint}), we eliminate the non-diagonal terms, i.e., the last two terms in the left hand side; we get then,  after some algebra 
\be
(2\del^+\del^--\D_{(0)}^2)\;A^i_{(1)}=D_{(0)}^i(\beta_{_0}-D_{(0)}^j\alpha^j_{_0}),
\ee
where $\D^2_{(0)}\equiv D_{(0)}^iD_{(0)}^i$.

In order to solve this equation, we introduce the Green's function $G_{_B}(x,y)$ that describes the propagation of a gluon  in the background field $A^i_{_B}$, that is, $G_{_B}(x,y)$ is solution of the equation
\be
(2\del^+\del^--\D^2_{(0)})\;G_{_B}(x,y)=\delta (x-y)\,,\labe{propB}
\ee
with retarded conditions ($G_{_B}$ is non-vanishing only when $x^\pm>y^\pm$).  The gauge field can then be expressed as follows (see Appendix \ref{Proof2} for details of the derivation)
\beq
&&A^i_{(1)}(x)=2\int\limits_{y^-,y^+=\epsilon} d^2\y \;G_{_B}(x,y)\;\alpha^i_{_0}(\y)\nn
&&+\int\limits_{y^+,y^->\epsilon} d^4 y\; G_{_B}(x,y)\left[D_{(0)}^i\left[\beta_{_0}(\y)-D_{(0)}^j\alpha^j_{_0}(\y)\right]\right].\labe{A1B1}
\eeq

It remains to solve Eq.~(\ref{propB}) for $G_{_B}(x,y)$. To this aim, we note that for any color vector $F$, we have 
\be
D_{(0)}^iF=V^\dag\del^i(VF)\;\;\text{and}\;\;\D^2_{(0)}F=V^\dag\bdel^2(VF).
\ee
Thus, with $\square\equiv 2\del^+\del^--\D_{(0)}^2$, we can write Eq.~(\ref{propB}) as
\be
V^\dag\square ( VG_{_B}(x,y))=\delta(x-y),\labe{propB2}
\ee
or \cite{AJMV}
\be\labe{propB3}
G_{_B}(x,y)=V^\dag(\x)G_{_0}(x-y)V(\y).
\ee
We used the fact that $V$ is independent of $x^+$ and $x^-$ in the forward light-cone, and $G_{_0}(x-y)$,  the free retarded propagator is known  \cite{BlaizGV1}:
\be
G_{_0}(x-y)=\theta(x^+-y^+)\theta(x^--y^-)\delta((x-y)^2).
\ee
The Wilson line $V$ attached to the two ends of the propagator   $G_{_B}$ in  Eq. (\ref{propB3}) reflects the presence of the pure gauge field $A^i_{_B}$ in the forward light cone. Would it be absent, $G_B$ would reduce simply to the free retarded propagator. Indeed we shall see shortly that a simple redefinition of the gauge condition allows us to eliminate $A^i_{_B}$, thereby simplifying the calculation.

Before doing so, it is useful to transform Eq.~(\ref{A1B1}) into a simpler equation. This is obtained by applying the operator $\square V$  on both sides of  Eq.~(\ref{A1B1}). Then, by using the fact that $\square (VG_{_B})=V$, one gets  
\be
\square (VA^i_{(1)})=2\delta(x^+)\delta(x^-)V\alpha^i_{_0}+\theta(x^+)\theta(x^-)\del^i(V\beta_{_0}-\del^j(V\alpha^j_{_0})),\labe{A1B2}
\ee   
where we have taken the limit $\epsilon \to 0$. We note that  the fields $A^i_{(1)}$, $\alpha^i_{_0}$ and $\beta_{_0}$   all appear multiplied by the same Wilson line $V$, which can therefore be absorbed in a redefinition of the fields: $\At^i_{(1)}=VA^i_{(1)}$, $\alphat^i_{_0}=V\alpha^i_{_0}$ and $\betat_{_0}=V\beta_{_0}$. The equation  for the gauge field $\At_{(1)}$ is then simply:
\be
\square \At^i_{(1)}=2\delta(x^+)\delta(x^-)\alphat^i_{_0}+\theta(x^+)\theta(x^-)\del^i(\betat_{_0}-\del^j\alphat^j_{_0}).\labe{A1B2bis}
\ee 
This redefinition of the fields involves the same gauge transformation as that which relates $G_B$ and $G_0$ in Eq.~(\ref{propB3}). This suggests that the calculation can indeed be made simpler by a proper redefinition of the  light-cone gauge condition. This is discussed in the next sub-section.

\subsection{The light-cone gauge $\del^-A^+=0$}\label{newgauge}

Consider then the  gauge rotation induced by the Wilson line $V$: 
\be
\At^\mu\cdot T=V(A^\mu\cdot T)V^\dag-\frac{1}{ig} V\del^\mu V^\dag.\label{gaugerotation3}
\ee
The effect of this gauge rotation on the field $A^\mu_{(0)}$ is given by
\beq\label{gaugerotation2}
\At_{(0)}^+\cdot T &=& -\frac{1}{ig} V\del^+ V^\dag=\Phi_{_B}\cdot T ,\nonumber\\
\At_{(0)}^- \cdot T &=&  V(A^-_{(0)}\cdot T )V^\dag-\frac{1}{ig} V\del^-V^\dag=V(\Phi_{_A}\cdot T )V^\dag, \nn
\At_{(0)}^i \cdot T &=& V(A^i_{(0)}\cdot T )V^\dag-\frac{1}{ig} V\del^i V^\dag=0.
\eeq
This gauge rotation eliminates the  transverse component  $A^i_{_B}$ and generates a non-zero $A^+$ component. Strictly speaking this gauge field is therefore no longer compatible with the light-cone gauge $A^+=0$. However, nothing prevents us to redefine the gauge condition so that $A^+= \Phi_{_B}$. Since $\Phi_{_B}$ exists only in a small strip $0\le x^-\le \epsilon$, this is equivalent to the condition $A^+=0$ almost everywhere, in particular in the forward light-cone. In fact, the slightly more general  definition of light-cone gauge, 
\be
\del^-A^+=0,\labe{LCG}
\ee
  includes all the gauge choices  that we have considered previously, including of course $A^+=0$ and $A^+= \Phi_{_B}$, and  enlarges the possibilities for fixing the initial fields.

Consider now the effect of the gauge rotation (\ref{gaugerotation3}) on the higher order terms of  the expansion (\ref{expan}). The inhomogeneous term in   Eq. (\ref{gaugerotation3}) is already taken into account in the transformation of $A^\mu_{(0)}$ into $\tilde A^\mu_{(0)}$, Eq.~(\ref{gaugerotation2}). Therefore,   for all $n>0$,   the fields $A^\mu_{(n)}$ transform homogeneously
\be
\At^\mu_{(n)}\cdot T =V(A^\mu_{(n)}\cdot T )V^\dag, 
\ee
or, by using the identity $V T^a V^\dagger= T^b V_{ba}$, 
\be
\At^\mu_{(n)}=VA^\mu_{(n)}\;.
\ee
This applies in particular to $A^\mu_{(1)}$, and this completes the proof that the redefinition of the fields from Eq. (\ref{A1B2}) to Eq. (\ref{A1B2bis}) is indeed a gauge rotation involving the Wilson line $V$.

\begin{figure}[ht]
\begin{center}
\resizebox*{!}{6cm}{\includegraphics{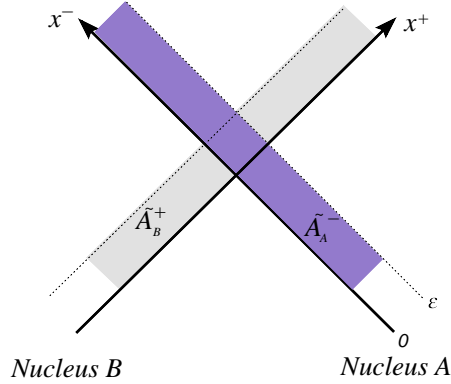}}
\caption{The gauge field before the collision  in the light-cone gauge $A^+=\Phi_{_B}$ is localized in a small strip of width $\epsilon$  along the light-cone.  }\label{fig3}
\end{center}
\end{figure}

After performing the gauge rotation (\ref{gaugerotation2}), which removes the pure gauge, the zeroth order field has the following components 
\be
\tilde A_{(0)}^+=\Phi_{_B},\qquad \tilde A_{(0)}^-=V\Phi_{_A},\qquad A_{(0)}^i=0, \labe{nip}
\ee
with the non vanishing components entirely localized in the small strip of width $\epsilon$ along the light cone (see Fig.~\ref{fig3}). 
The  gauge field  at the border of this strip are then simply those of Eqs.~(\ref{alpha0}) and (\ref{beta0}), to within the gauge rotation induced by $V$:
\beq\labe{alpha0VU}
&&\alphat_{_0}^i=V\alpha_{_0}^i=V(U-1)A^i_{_B},\\
&&\betat_{_0}=V\beta_{_0}=(\del^j U)A^j_{_B},\labe{beta0VU}
\eeq
and are  produced by the following conserved current (see Eq.~(\ref{Jmu}))
\beq\label{JmuVU}
{\tilde J}^\mu=\delta^{\mu-}V\rho_{_A}+\delta^{\mu+}VU V^\dag \rho_{_B}.
\eeq

The particular choice of gauge discussed in this subsection makes the calculation of gluon production easier and physically more transparent since all the pure gauge fields  are eliminated in the forward light-cone .

\section{Gluon production\label{gluprod}}


We now exploit the simplification brought by the use of the modified gauge condition $A^+=\Phi_{_B}$, in  calculating the spectrum of gluons produced just after the collision. As argued earlier only the leading correction in the expansion  (\ref{expan}) is needed in this calculation, that is, $\At^\mu_{(1)}$. Let us recall that the  zeroth order $\At^\mu_{(0)}$ is given by Eq.~(\ref{nip}). 
Since in the forward light cone $x^\pm>\epsilon$, $\Phi_{_A}=\Phi_{_B}=0$,  we have ${\tilde A}^\mu_{(0)}=0$. It follows that in the forward light cone, $D_{(0)}^i=\partial^i$, and the  linearized Yang-Mills equations take the simple form:
\beq
&&\del^+(\del_\mu\At^\mu_{(1)})=0,\labe{YMtlin+}\\
&& \square \At^-_{(1)}-\del^-\del_\mu\At^\mu_{(1)}=0,\labe{YMtlin-}\\
&&\square \At^i_{(1)}-\del^i\del_\mu\At^\mu_{(1)}=0.\labe{YMtlint}
\eeq
Using the same technique as in the previous section to handle the boundary conditions, we get from them the following equations for the non vanishing components of the gauge field:
\beq
&&\square \At^i_{(1)}=2 \delta(x^+)\delta(x^-)\alphat^i_{_0}+\theta(x^+)\theta(x^-)\del^i(\betat_{_0}-\del^j\alphat^j_{_0}),\nn
&&\square \At^-_{(1)}=2 \delta(x^+)\theta(x^-)\betat_{_0}.\labe{res}
\eeq
As expected, the first equation (\ref{res}) is identical to Eq. (\ref{A1B2bis}). As we shall see, this is in fact  the only one needed to calculate gluon production.

The spectrum of produced gluon is given by (see e.g. \cite{BlaizGV1})
\be
(2\pi)^32E\frac{dN}{d^3{\bf q}}=\sum\limits_\lambda \langle |{\cal M}_\lambda|^2\rangle,
\ee
where $\bf q$ is the three-momentum of the gluon, $E=|{\bf q}|$ its energy, $\lambda$ its polarization. The symbol $\langle...\rangle$ stands for the average over the color sources $\rho_{_A}$ and  $\rho_{_B}$. The amplitude ${\cal M}_\lambda$ is related to the classical gauge field by
\be
{\cal M}_\lambda=\lim_{q^2\rightarrow 0} q^2\At^i(q)\epsilon^i_\lambda(q) ,\labe{amp}
\ee
where $ \epsilon^i_\lambda(q)$ is the  polarization vector of the gluon and $q$ its four-momentum. In the axial gauge $A^+=0$, only the transverse components of the field contribute, and the sum over polarizations states is done with the help of the relation $\sum\limits_\lambda \epsilon^i_\lambda(q)\epsilon^{\ast j}_\lambda(q)=\delta^{ij}$.

It is easy to see that only $\At^i_{(1)}(q)$ contribute to gluon production ($\At^i_{(0)}(q)$ vanishes in the forward light-cone).  In Fourier space Eqs.~(\ref{res}) read
\beq
&&-q^2\At^i_{(1)}(q)=-2 \left(\delta^{ij}-\frac{q^iq^j}{2q^+q^-}\right)\alphat^j_{_0}(\q)-2i\frac{q^i}{2q^+q^-}\betat_{_0}(\q),\nn
&&-q^2\At^-_{(1)}(q)=-\frac{2i}{q^+}\betat_{_0}(\q),\labe{res2}
\eeq
with
\beq
 \alphat^j_{_0}(\q)=\int d^2\x \;e^{-i\q\cdot\x}\alphat^j_{_0}(\x),
\eeq
and similarly for $\betat_0(\q)$. 
Note that, for on-shell gluons ($2q^+q^--\q^2=0$), the condition of transversality $q_\mu {\cal M}^\mu(q)=0$ is fulfilled, where ${\cal M}^\mu\equiv \lim_{q^2\rightarrow 0} q^2\At^\mu(q)$.
By inserting  in  (\ref{amp}) the explicit expression of $\At^i_{(1)}(q)$ given in Eq.~(\ref{res2})  we get
\be
{\cal M}_\lambda=-2 \left(\epsilon_\lambda^j-\frac{\epsilon_\lambda\cdot\q}{\q^2}q^j\right)\alphat^j_{_0}(\q)-2i\frac{\epsilon_\lambda\cdot\q}{\q^2}\betat_{_0}(\q).\labe{res3}
\ee
At this point we make the following  decomposition of $\alpha^i_{_0}$
\be
\alphat^i_{_0}=\frac{q^i}{\q^2}(\q\cdot\balphat_{_0}(\q))+\varepsilon^{ij}\frac{q^j}{\q^2}(\q\times \balphat_{_0}(\q)),\labe{deco}
\ee
where $\varepsilon^{ij}$ is the antisymmetric tensor with $\varepsilon^{12}=1$. This   allows us to write the gluon spectrum in the following compact form:
\be
4\pi^3E\frac{dN}{d^3{\bf q}}=\frac{1}{\;\q^2}\langle |\q\times \balphat_{_0}(\q)|^2 +|\betat_{_0}(\q)|^2\rangle.\labe{resN}
\ee

The formula (\ref{resN}) is the main result of this paper. It provides a compact expression for the spectrum of gluons produced in a nucleus-nucleus collision in terms of the fields $\tilde\alpha_{_0}$ and $\tilde\beta_{_0}$ that exist on the light-cone just after the collision. As shown by their explicit expressions (\ref{alpha0VU}) and (\ref{beta0VU}), these fields are determined entirely by the functions $\Phi_{_A}$ and $\Phi_{_B}$ that are obtained as a function of the sources $\rho_{_A}$ and $\rho_{_B}$ by solving Eqs.~(\ref{Phi_A})  and (\ref{Phi_B}). The complex color structure of the product of the Wilson lines $U$ and $V$ hidden in $\tilde\alpha_{_0}$ and $\tilde\beta_{_0}$ prevents one to write Eq.~(\ref{resN}) in a $k_t$-factorized form. This property of  $k_t$-factorization is recovered, however, when one of the sources is weak, as we shall see shortly.  The spectrum (\ref{resN}) is rederived   in Appendix \ref{FSG} using a variant of the Fock-Schwinger gauge which makes the calculation more symmetrical: this provides an indication of the robustness of the calculation against variations in the gauge choice. 

The formula  (\ref{resN}) is obviously not exact since it is the leading order of an expansion (it involves only $\At^i_{(1)}$). However the initial values of the fields on the light-cone are calculated exactly, i.e., taking the full non-linearity of the Yang-Mills equations into account. Such non-linearities are ignored in the calculation of the field $\At_{(1)}$ which propagates  in the forward light-cone without further interactions. The next order in the expansion (\ref{expan}), ${\tilde A}_{(2)}$, describes in particular the merging  of  such gluons. Note that such mergings are taking place  at later times since $\At_{(2)}=0$ when $x^\pm=\epsilon$.

The formula  (\ref{resN}) becomes exact  when one of the nuclei can be considered as a weak source. To see that let us expand the Wilson $U$ line at leading order in $\rho_{_A}$ 
\be
U(x)\equiv1+ig\int_0^\epsilon dx^+ \Phi_{_A}(x^+,\x)\cdot T,
\ee
which allows us to write 
\beq
\alphat^i_{_0}(\x)=igVT^aA^i_{_B}\Phi^a_{_A}(\x)
               =(\del^iV)\Phi_{_A}(\x),
\eeq
and
\beq
\betat_{_0}(\x)=igVT^aA^j_{_B}\del^j\Phi^a_{_A}(\x) 
              = (\del^jV)\del^j\Phi_{_A}(\x),
\eeq
where we have used the identity
\be
(VT^aA^i_{_B})_b=\frac{1}{ig}(\del^iV)_{ba},
\ee
and we have set $\int_0^\epsilon dx^+ \Phi_{_A}(x^+,\x)=\Phi_{_A}(\x)$.
In Fourier space, setting $\p=\q-\k$, we have
\beq\labe{alphaFT}
&&\alphat_{_0}^i(\q)=-i\int \frac{d^2 \k}{(2\pi)^2} \p^i \;V(\p)\;\Phi_{_A}(\k),\\ 
&&\betat_{_0}(\q)=-\int \frac{d^2 \k}{(2\pi)^2}(\p\cdot\k)\;V(\p)\;\Phi_{_A}(\k),\labe{betaFT}
\eeq
with
\be
 V(\p)=\int d^2\x \;e^{-i\p\cdot\x}(V(\x)-1).
\ee
By replacing in (\ref{resN})  $\alphat_{_0}^i(\q)$ and $\betat_{_0}(\q)$ by their  expressions (\ref{alphaFT}) and (\ref{betaFT})  one recovers  the well known formula for proton-nucleus collisions  (see e.g. Eq.~(22) of Ref.~\cite{GM}):
\be
4\pi^3E\frac{dN}{d^3{\bf q}}=\frac{1}{\q^2}\int \frac{d^2 \k}{(2\pi)^2} \frac{d^2 \k'}{(2\pi)^2}\;{\cal K}(\k,\k',\q)\;\langle V^\dag(\p')V(\p)\rangle_{ab}\;\langle \Phi_{_A}^{a\ast}(\k')\Phi_{_A}^{b}(\k)\rangle ,\label{resNpA1}
\ee
where $\p'=\q-\k'$ and ${\cal K}(\k,\k',\q)$ is  the square of the Lipatov vertex 
\beq
&&{\cal K}(\k,\k',\q)=(\p\times\k)(\p'\times\k')+(\p\cdot\k)(\p'\cdot\k').
\eeq
This formula exhibits $k_t-$factorization.

Although the formalism that we are using is not manifestly symmetric (because of the gauge choice), it is easy to verify that a similar result can be obtained in the case where the  weak source is  $\rho_{_B}$:
\beq
&&\tilde \alpha_{_0}^i(\q)=-i\int \frac{d^2 \k}{(2\pi)^2} \p^i \;U(\k)\;\Phi_{_B}(\p),\\
&&\tilde \beta_{_0}(\q)=-\int \frac{d^2 \k}{(2\pi)^2}(\p\cdot\k)\;U(\k)\;\Phi_{_B}(\p),
\eeq
which leads to
\be
4\pi^3E\frac{dN}{d^3{\bf q}}=\frac{1}{\q^2}\int \frac{d^2 \k}{(2\pi)^2} \frac{d^2 \k'}{(2\pi)^2}\;{\cal K}(\k,\k',\q)\;\langle U^\dag(\k')U(\k)\rangle_{ab}\;\langle \Phi_{_B}^{a\ast}(\p')\Phi_{_B}^{b}(\p)\rangle .\labe{resNpA2}
\ee
This is identical to Eq. (\ref{resNpA1}) after the substitutions $A\to B$,  $V\to U$ and $\p\leftrightarrow\k$, $\p'\leftrightarrow\k'$, under which the kernel ${\cal K}$ is invariant.


\section{Conclusion and perspectives}


We have shown that the calculation of gluon production in nucleus-nucleus collisions in the color glass framework can be greatly simplified by an appropriate choice of gauge. The particular  axial gauge that we have proposed, $\partial^-A^+=0$ (or more precisely $A^+=\Phi_{_B}$), allows us to determine explicitly the conserved current, and the fields immediately after the collisions, in terms of the color sources of the two nuclei, given in the covariant gauge. The fields before the collision are localized in an infinitesimal strip along the light cone. In particular, no pure gauge fields are left in the forward light cone; in other gauges the presence of such pure gauges complicates the calculations and sometimes obscures the physical interpretation. 

We have obtained an explicit expression, Eq.~(\ref{resN}),  for the spectrum of gluons produced in nucleus-nucleus collisions, which holds in leading order in a new expansion where the non linearities of the Yang-Mills equations in the forward light-cone are treated iteratively. The leading order is obtained by linearizing the Yang-Mills equations in the forward light cone, keeping as boundary values for the fields, the exact  fields that exist on the light-cone immediately after the collision. Higher order terms in the expansion describe ``final state interactions'',  in particular the merging of gluons; these, however, take place at later times, so that our formula exactly calculates the spectrum of gluons produced at a short time after the collision. It would be  interesting to investigate quantitatively  the importance of the  final state interactions, for instance by  comparing (\ref{resN}) to available numerical studies for gluon production \cite{KrasnNV,Lappi1} and for the energy density \cite{KrasnNV,Lappi2}, and check whether these are negligible, as argued in Ref. \cite{KovAA}.

The iterative solution that we propose is not an expansion in an obvious small parameter. However it lends itself to simple calculations whenever such a parameter is  present. Thus, for instance, the formula (\ref{resN}) provides the exact leading order result when expanded in powers of the color source of one of the two nuclei. An immediate application of the present formalism is to obtain the next-to-leading order in this expansion. This requires the determination of the term $A^i_{(2)}$ in our expansion. The result of this calculation will be presented in a forthcoming publication.

\appendix

\section{$A^-$ along the $x^+$ axis}\label{Proof1}


In this appendix we establish the continuity of  $A^-(x^-)$ at $x^-=0$, that is, we show that 
\be
\lim_{\epsilon\rightarrow 0} A^-(x^+,x^-=\epsilon,\x)=\Phi_{_A}(x^+,\x).
\ee
To do so, let us rewrite the Yang-Mills  equation (\ref{YM1}) in the form
\be
\del^+(\del^+ A^-)=\del^+(\del^i A^i)-J^+-ig(A^i\cdot T)\;\del^+ A^i,
\ee
and integrate it over $x^-$ from $-\infty$ to $x^-$. Using the initial conditions $A^\mu(x^-<0,x^+,\x)=\delta^{\mu-}\Phi_{_A}(x^+,\x)$ and $J^+(x^-<0)=0$, we obtain
\be
\del^+ A^-(x^-)=\del^i A^i(x^-)-\int^{x^-}_0dy^- \left[   J^+(y^-)+ig(A^i\cdot T)\;\del^+ A^i(y^-)\right],\label{int1const}
\ee
where the dependence on $x^+$ and $\x$ has been omitted to alleviate the notations.  
After a further integration of Eq.~(\ref{int1const}) over $x^-$,  from $0$ to $\epsilon$,  one gets
\beq
\!\!\!\!\!\!\!\!\!\!A^-(\epsilon)-\Phi_{_A}&\!\!=\!\!&\int^{\epsilon}_0 dx^-\del^i A^i(x^-)\nn &\!\!-\!\!&\int^{\epsilon}_0 dx^- \int^{x^-}_0\!\!dy^-  \left[   J^+(y^-)\!+\!ig(A^i\cdot T)\;\del^+ A^i(y^-)\right]\!.\label{int2const}
\eeq
 The continuity of $A^-(x^-)$ follows then from the absence, in the limit $\epsilon\to 0$, of $\delta$-function singularities in the integrands in the right hand side (such   singularities occur  in the current $J^+$, or in $\del^+ A^i$, but these disappear after the first integration,  over $y^-$). 

\section{Proof of equation (\ref{A1B1})}\label{Proof2}

We are looking for a solution of an equation of the form 
\be
(2\del^+\del^--\D_{(0)}^2)\;A^i(y)=J^i(y),\labe{EquaA1}
\ee
in the region $x^+\geq\epsilon$ and $x^-\geq \epsilon$, which we shall obtain by using the retarded Green's function  $G_{_B}$ defined in Eq.~(\ref{propB}). Let us multiply Eq. (\ref{EquaA1}) by $G_{_B}$  and integrate over $y$ in the forward light cone:
\be
\int\limits_{y^+,y^->\epsilon}  d^4y  \;G_{_B}(x,y)(2\del^+\del^--\D_{(0)}^2)\;A^i(y)=\int\limits_{y^+,y^->\epsilon}  d^4y \;G_{_B}(x,y)J^i(y),\labe{EquaA2}
\ee
Loosely speaking, the Green's function $G_{_B}$ is the inverse of the differential operator which sits next to it in Eq.~(\ref{EquaA2}). However, care must be exerted in using this property because the $y$-integration covers only the forward light-cone and, as we shall see, boundary terms will survive integrations by parts.

Let us first consider the first term in the left hand side of Eq. (\ref{EquaA2}). Performing successive partial integrations we end up with 
\beq
&&\int\limits_{y^+,y^->\epsilon}  d^4y \;G_{_B}(x,y)\stackrel{\rightarrow}{\del^+_y}\stackrel{\rightarrow}{\del^-_y}A^i(y)=\int\limits_{y^+,y^->\epsilon}  d^4y \;G_{_B}(x,y)\stackrel{\leftarrow}{\del^+_y}\stackrel{\leftarrow}{\del^-_y}A^i(y)\;\nn
&&+\int\limits_{y^-=\epsilon} dy^+d^2\y \;G_{_B}(x,y)\stackrel{\leftarrow}{\del^-_y}A^i(y)+\int\limits_{y^+=\epsilon} dy^-d^2\y\; G_{_B}(x,y)\stackrel{\leftarrow}{\del^+_y}A^i(y)\nn
&&+\int\limits_{y^+,y^-=\epsilon} d^2\y \;G_{_B}(x,y)\;A^i(y),
\eeq
Since $A^i(y^+=\epsilon,y^-\geq\epsilon,\y)=A^i(y^+\geq\epsilon,y^-=\epsilon,\y)=A^i(\y)$,  as shown in Sect. \ref{fieldborder}, the integrals over $y^+$ and $y^-$ in the second and third term in the right hand side can be performed explicitly,  leading to 
\beq\label{equaB120}
&&\int\limits_{y^+,y^->\epsilon}  d^4y \;G_{_B}(x,y)\stackrel{\rightarrow}{\del^+_y}\stackrel{\rightarrow}{\del^-_y}A^i(y)=\int\limits_{y^+,y^->\epsilon}  d^4y \;G_{_B}(x,y)\stackrel{\leftarrow}{\del^+_y}\stackrel{\leftarrow}{\del^-_y}A^i(y)\;\nn
&&\qquad\qquad-\int\limits_{y^+,y^-=\epsilon} d^2\y \;G_{_B}(x,y)\;A^i(\y).
\eeq
We turn  now to the second term in the left hand side of (\ref{EquaA2}). By performing successive partial integrations, and assuming that the field $A^i(x)$ vanishes when $|\x|\rightarrow \infty$ we get 
\be\label{equaB121}
\int\limits_{y^+,y^->\epsilon}  d^4y \;G_{_B}(x,y)\stackrel{\rightarrow}{\D}^2_{(0)y}\;A^i(y)=\int\limits_{y^-,y^+>\epsilon}  d^4y\; G_{_B}(x,y)\stackrel{\leftarrow}{\D^\dag}^2_{(0)y}\;A^i(y),
\ee
where $D^{\dag\;i}_{(0)}=\del^i+igA^i_{_B}\cdot T$.
Finally, combining Eqs.~(\ref{equaB120}) and (\ref{equaB121}), we obtain
\beq
&&\int\limits_{y^+,y^->\epsilon} d^4y \;G_{_B}(x,y)(2\stackrel{\leftarrow}{\del_y^+}\stackrel{\leftarrow}{\del_y^-}-\stackrel{\leftarrow}{\D^\dag}^2_{(0)y})\;A^i(y)=A^i(x)\nn
&& =2\int\limits_{y^+,y^-=\epsilon} d^2\y \;G_{_B}(x,y)\;A^i(\y)+\int\limits_{y^+,y^->\epsilon} d^4y\;G_{_B}(x,y)\;J^i(y),\nonumber\\
\eeq
from which Eq. (\ref{A1B1}) follows, with  $J^i=D_{(0)}^i(\beta_{_0}-D_{(0)}^j\alpha^j_{_0})$.


\section{Comparison  with the Fock-Schwinger gauge}\label{FSG}

In this appendix we briefly review the calculation of the gauge field in the axial gauge 
 \beq\label{FSgauge}
 x^+{\cal A}^-+x^-{\cal A}^+=0,
 \eeq 
often referred to as the Fock-Schwinger gauge, and compare with the results obtained in this paper.
 In the gauge (\ref{FSgauge}), we have ${\cal A}^-=0$ along the $x^+$ axis, and ${\cal A}^+=0$ along the $x^-$ axis.
 It is then easy to verify that the current 
 \beq\label{FScurrent}
 J^+ \equiv V^\dag \rho_{_B}, \qquad J^-\equiv U^\dag \rho_{_A},\qquad J^i =0\;, 
 \eeq
 is covariantly conserved, and  that the gauge field components  before the collision  are transverse:
\beq
&&{\cal A}^i={\cal A}^i_{_A}\qquad\text{for}\;\; x^+>0,\nn
&&{\cal A}^i={\cal A}^i_{_B}\qquad \text{for}\;\; x^->0,\nn
&&{\cal A}^+={\cal A}^-=0,
\eeq
where 
 \beq\label{FSfield}
 {\cal A}^i_{_A} \cdot T=-\frac{1}{ig}U^\dagger  \partial^i U , \qquad {\cal A}^i_{_B} \cdot T=-\frac{1}{ig}V^\dagger  \partial^i V ,
 \eeq
 with $U=U(x^+,\x)$ given by Eq.~(\ref{WilsonU1}) and  $V=V(x^-,\x)$ by Eq.~(\ref{WilsonV1}).

The gauge choice (\ref{FSgauge}) suggests the following ansatz for the components ${\cal A}^{\pm}(x)$ of the gauge field after the collisions \cite{JKMW1}:
\be\labe{FSform}
{\cal A}^+(x)=-x^+\beta_{_{FS}}(\tau,\x),\qquad {\cal A}^-(x)=x^-\beta_{_{FS}}(\tau,\x), 
\ee
where $\tau=2\sqrt{x^+x^-}$ is the proper time. The gauge field  $A(x)$ in the light-cone gauge is related to the gauge field ${\cal A}(x)$ of  the Fock-Schwinger gauge  by a gauge rotation:
\beq\label{gaugerotation4}
A^+\cdot T &=&  \Omega^\dag ({\cal A}^+\cdot T)\, \Omega-\frac{1}{ig} \Omega^\dag\del^+ \Omega=0,\nonumber\\
A^- \cdot T&=&  \Omega^\dag ({\cal A}^-\cdot T)\,\Omega-\frac{1}{ig} \Omega^\dag\del^-\Omega,\nn
A^i\cdot T &=& \Omega^\dag ({\cal A}^i\cdot T)\,\Omega-\frac{1}{ig} \Omega^\dag\del^i\Omega.
\eeq
Is is easy to verify that, given the ansatz (\ref{FSform}), there exists a boost invariant solution of the equations above, that is, a function $\Omega$ that depends only on the proper time $\tau$, $\Omega(x^+,x^-,\x)= \Omega(\tau,\x) $.  To show this, we divide the first equation by $x^+ $ and use Eq.~(\ref{FSform}) in order  to express ${\cal A}^+$  in terms of $\beta_{_{FS}}$. Assuming then that $\Omega$ depends only on $\tau$, we obtain 
\beq\label{gaugerotation5}
4\frac{\del}{\del\tau^2} \Omega(\tau,\x)+ig(\beta_{_{FS}}(\tau,\x)\cdot T)\;\Omega(\tau,\x)=0,
\eeq
whose solution reads
\be\label{WilsonUFS} 
\Omega(\tau,\x)\equiv {\mathcal
P}\exp{\left[-\frac{ig}{4}\int_{0}^{\tau^2}d\xi^2\beta_{_{FS}}(\xi,\x)\cdot T\right]}.
\ee 
By dividing the second equation  (\ref{gaugerotation4}) by $x^-$ and assuming for $A^-(x)$ a form analogous to that  of ${\cal A}^-(x)$ in Eq.~(\ref{FSform}), i.e., $A^-(x)=x^-\beta_{_{LC}}(\tau,\x)$, we obtain 
\beq\label{gaugerotation6}
\beta_{_{LC}}\cdot T&=& \Omega^\dag(\beta_{_{FS}}\cdot T)\,\Omega-\frac{4}{ig} \Omega^\dag\frac{\del}{\del\tau^2}\Omega\nn
     &=& \Omega^\dag(\beta_{_{FS}}\cdot T)\,\Omega-\frac{1}{ig} \Omega^\dag(-ig\beta_{_{FS}}\cdot T)\,\Omega\nn
     &=&2\;\Omega^\dag(\beta_{_{FS}}\cdot T)\,\Omega.
\eeq
Near the light-cone, i.e., when $\tau\to 0$, $\Omega\rightarrow 1$. Then, from  (\ref{gaugerotation4})  and (\ref{gaugerotation6}) we get, respectively, 
\be
\Ac^i(0,\x)=A^i(0,\x), \qquad \beta_{_{FS}}(0,\x)=\frac{1}{2}\beta_{_{LC}}(0,\x),\labe{betaLC-FS}
\ee
where we have assumed that the transverse components of the gauge field, $\Ac^i$ and $A^i$, depend only on $\tau$ (with $\Ac(0,\x)\equiv \Ac(\tau=0,\x)$ and similarly for $A(0,\x)$).

In the light-cone gauge, with the further choice that leads to the same initial (transverse) fields as in Eq.~(\ref{FSfield}), i.e., $A_{_A}^i=\Ac_{_A}^i$ and $A_{_B}^i=\Ac_{_B}^i$, the gauge field on the light-cone is given by Eq. (\ref{Aborder2}):
 \beq
 \beta_{_{LC}} =\del^+A^- =(\Ac^i_{_A} \cdot T) \Ac^i_{_B} ,\qquad A^i =\Ac^i_{_A} +\Ac^i_{_B} . \eeq
 Therefore, we reproduce the well known result for the gauge field on the light-cone in the Fock-Schwinger gauge, namely \cite{JKMW1,JKMW2,GV}\footnote{In the references quoted here the fields are expressed in matrix representation, i.e. ${\cal A}\cdot T\to {\cal A}$, and $\beta_{_{FS}}=\frac{1}{2}[ {\cal A}^i_{_A}, {\cal A}^i_{_B}]$}:
\beq\label{FSborder}
&&{\cal A}^i(0,\x)={\cal A}^i_{_A}(\x)+{\cal A}^i_{_B}(\x),\nn
&&\del^+{\cal A}^-(0,\x)=-\del^-{\cal A}^+(0,\x)=\beta_{_{FS}}(0,\x)=\frac{1}{2}({\cal A}^i_{_A}(\x)\cdot T) {\cal A}^i_{_B}(\x).\nn
\eeq
At this point we depart from the traditional treatments in the Fock-Schwinger gauge. 
In complete analogy with what we did in Sect. \ref{LCGfield},   we eliminate  the  transverse pure gauge fields   by performing a gauge rotation involving the products  $UV$ or $VU$ of the Wilson lines (\ref{WilsonU1}) and (\ref{WilsonV1})\footnote{The elimination of the pure gauge components of the gauge fields is useful in view of the calculation of gluon production, as was also observed in  \cite{DumitM1}.}. By doing so, we generate a longitudinal initial field (for $t<0$)
\be\labe{FSinfield}
{\cal A}^+=\Phi_{_B},\qquad {\cal A}^-=\Phi_{_A},\qquad {\cal A}^i=0.
\ee
Such a field is   no longer compatible with the gauge fixing condition (\ref{FSgauge}). Consider however the following extension of this condition, 
\be\label{FSgauge2}
\del^-{\cal A}^++\del^+{\cal A}^-=0,
\ee
which may be viewed as the symmetric version of the gauge condition (\ref{LCG}).
The initial field (\ref{FSinfield}) clearly fulfills this condition. Besides, in the forward light-cone, and provided the ansatz (\ref{FSform}) is verified,   the gauge conditions (\ref{FSgauge2})  and   (\ref{FSgauge}) are simultaneously verified.

We shall use now this symmetric gauge condition to calculate  gluon production in leading order in the expansion (\ref{expan}), and recover the results of Sect.~\ref{gluprod}. Following the steps described in Sect.~\ref{gluprod}, one obtains first the linearized Yang-Mills equations in the forward light-cone (compare with Eqs.~(\ref{YMtlin+}), (\ref{YMtlin-}), (\ref{YMtlint})):
\beq\labe{FSYM}
&&\square {\cal A}^+_{(1)}=-\del^+\!\del^j{\cal A}^j_{(1)},\nn
&&\square {\cal A}^-_{(1)}=-\del^-\!\del^j{\cal A}^j_{(1)},\nn
&&\square {\cal A}^i_{(1)}=-\del^i \del^j{\cal A}^j_{(1)},
\eeq
with ${\cal A}_{(1)}(\tau=0,\x)={\cal A}(\tau=0,\x)$ and ${\cal A}_{(n)}(\tau=0,\x)=0$ for $n>1$. The values of the fields, created by the initial fields (\ref{FSinfield}) near the light-cone, are related to those in light-cone gauge (given in  Eqs. (\ref{alpha0VU}) and (\ref{beta0VU})), through Eq. (\ref{betaLC-FS}). More explicitly, ${\cal A}^i(\tau=0,\x)=\alphat^i_{_0}(\x)$ and $\del^+{\cal A}^-(\tau=0,\x)=-\del^-{\cal A}^+(\tau=0,\x)=\frac{1}{2}\betat_{_0}(\x)$.
The first two equations in (\ref{FSYM}), together with the gauge condition (\ref{FSgauge2}), lead to
\be
\square \left[ \del^+{\cal A}^-_{(1)}+\del^-{\cal A}^+_{(1)}\right]=2\del^+\del^-\del^j{\cal A}^j_{(1)}=0, 
\ee
from which it follows that the divergence of the transverse field is conserved, i.e.
\be
\del^j{\cal A}^j_{(1)}=\del^j\alphat_{_0}^j.
\ee
This allows us to rewrite the Yang-Mills equations in  a form that makes the boundary conditions explicit. Following the same steps as in Sect.~\ref{forwardLC}, we obtain \beq
&&\square {\cal A}^i_{(1)}=2\delta(x^+)\delta(x^-)\alphat_{_0}^i-\theta(x^+)\theta(x^-)\del^i(\del^j\alphat_{_0}^j),\\
&&\square {\cal A}^+_{(1)}=\delta(x^-)\theta(x^+)\betat_{_0},\\
&&\square {\cal A}^-_{(1)}=\delta(x^+)\theta(x^-)\betat_{_0}.
\eeq
In Fourier space we get
\be
-q^2{\cal A}^\mu_{(1)}(q)=-2\left(\delta^{ij}-\frac{q^iq^j}{2q^+q^-}\right)\alphat_{_0}^j(\q)\;\delta^{\mu i}-\frac{i}{q^-}\betat_{_0}(\q)\;\delta^{\mu+}-\frac{i}{q^+}\betat_{_0}(\q)\;\delta^{\mu-}.
\ee
From here, it is straightforward to get the gluon spectrum using the decomposition (\ref{deco}):
\be
4\pi^3E\frac{dN}{d^3{\bf q}}=\frac{1}{\q^2}\langle |\q\times \balphat_{_0}(\q)|^2 +|\betat_{_0}(\q)|^2\rangle,
\ee
which is identical to  Eq. (\ref{resN}) obtained in  light-cone gauge.

\section{Acknowledgments}

We would like to thank  D. Dietrich, F. Gelis, Yu. V. Kovchegov, R. Venugopalan for helpful discussions.

\bibliographystyle{unsrt}

\begin{thebibliography}{10}

\bibitem{McLerV}
{L.D. McLerran, R. Venugopalan}, Phys. Rev. {\bf D} {\bf 49}, 2233 (1994);   Phys. Rev. {\bf D} {\bf 49}, 3352 (1994); Phys. Rev. {\bf D} {\bf 50}, 2225 (1994).

\bibitem{JalilKLW}
{J. Jalilian-Marian, A. Kovner, A. Leonidov, H. Weigert}, Nucl. Phys. {\bf B}
  {\bf 504}, 415 (1997); Phys. Rev. {\bf D}
  {\bf 59}, 014014 (1999);  Phys. Rev. {\bf D}
  {\bf 59}, 034007 (1999);  Erratum. Phys. Rev.
  {\bf D} {\bf 59}, 099903 (1999).

\bibitem{KovneM1}
{A. Kovner, G. Milhano}, Phys. Rev. {\bf D} {\bf 61}, 014012 (2000).

\bibitem{KovneMW3}
{A. Kovner, G. Milhano, H. Weigert}, Phys. Rev. {\bf D} {\bf 62}, 114005
  (2000).

\bibitem{JalilKMW1}
{J. Jalilian-Marian, A. Kovner, L.D. McLerran, H. Weigert}, Phys. Rev. {\bf D}
  {\bf 55}, 5414 (1997).
\bibitem{IancuLM}
{E. Iancu, A. Leonidov, L.D. McLerran}, Nucl. Phys. {\bf A} {\bf 692}, 583
  (2001); Phys. Lett. {\bf B} {\bf 510}, 133
  (2001).

\bibitem{FerreILM1}
{E. Ferreiro, E. Iancu, A. Leonidov, L.D. McLerran}, Nucl. Phys. {\bf A} {\bf
  703}, 489 (2002).

\bibitem{Balit1}
{I. Balitsky}, Nucl. Phys. {\bf B} {\bf 463}, 99 (1996).
\bibitem{Kovch3}
{Yu.V. Kovchegov}, Phys. Rev. {\bf D} {\bf 60}, 034008 (1999).

\bibitem{cargese}
{E. Iancu, A. Leonidov, L.D. McLerran}, Lectures given at  the Cargese Summer School  ``QCD perspectives on hot and dense matter'', Cargese, France, 6-18 August 2001, hep-ph/022270. 

\bibitem{qgp3}
{E. Iancu and R. Venugopalan}, in R.C. Hwa, X.N. Wand (eds.), Quark-Gluon Plasma 3, World Scientific, Singapore 2003, hep-ph/0303204.


\bibitem{KrasnV}
{A. Krasnitz, R. Venugopalan}, Phys. Rev. Lett. {\bf 84}, 4309 (2000); Phys. Rev. Lett. {\bf 86}, 1717 (2001).

\bibitem{KrasnNV}
{A. Krasnitz, Y. Nara, R. Venugopalan}, Nucl. Phys. {\bf A} {\bf 727}, 427
  (2003);   Phys. Rev. Lett. {\bf 87}, 192302
  (2001).

\bibitem{Lappi1}
{T. Lappi}, Phys. Rev. {\bf C} {\bf 67}, 054903 (2003).


\bibitem{JKMW1}
A. Kovner, L.D. McLerran and H. Weigert, Phys. Rev. {\bf D 52}, 6231- (1995).

\bibitem{JKMW2}
J. Jaalilian-Marian, A. Kovner, L.D. McLerran and H. Weigert, Phys. Rev. {\bf D 55}, 5414 (1997).



\bibitem{GV}
F. Gelis and R. Venugopalan, ``Lectures on multi-particle production in the Glasma", hep-ph/0611157.


\bibitem{KovAA}
Yu. V. Kovchegov, Nucl. Phys. {\bf A} {\bf 692}, 557 (2001).


\bibitem{Balit2}
{I. Balitsky}, Phys. Rev. {\bf D} {\bf 70}, 114030 (2004).


\bibitem{KovchM3}
{Yu.V. Kovchegov, A.H. Mueller}, Nucl. Phys. {\bf B} {\bf 529}, 451 (1998).

\bibitem{KovnW}
{A. Kovner, U.A. Wiedemann}, Phys. Rev. {\bf D} {\bf 64}, 114002 (2001).

\bibitem{KovchT1}
{Yu.V. Kovchegov, K. Tuchin}, Phys. Rev. {\bf D} {\bf 65}, 074026 (2002).

\bibitem{DumitM1}
{A. Dumitru, L.D. McLerran}, Nucl. Phys. {\bf A} {\bf 700}, 492 (2002).

\bibitem{BlaizGV1}
{J.P. Blaizot, F. Gelis, R. Venugopalan}, Nucl. Phys. {\bf A} {\bf 743}, 13
  (2004).


\bibitem{GM}
F. Gelis and Y. Mehtar-Tani, Phys. Rev. {\bf D 73}, 034019 (2006).


\bibitem{Lappi:2007ku}
  T.~Lappi,
  Eur.\ Phys.\ J.\  C {\bf 55}, 285 (2008)
  [arXiv:0711.3039 [hep-ph]].

\bibitem{Fukushima:2007ki}
  K.~Fukushima,
  Phys.\ Rev.\  D {\bf 77}, 074005 (2008)
  [arXiv:0711.2364 [hep-ph]].

\bibitem{AJMV}
{A.Ayala, J. Jalilian-Marian, L. D. McLerran, R. Venugopalan}, Phys. Rev. {\bf D} {\bf 52},2935 (1995).

\bibitem{Fries}
R.J. Fries, J.I. Kapusta, Y. Li, nucl-th/064054.

\bibitem{Lappi2}
T. Lappi, Phys. Lett. {\bf B} {\bf 643} 11 (2006). 


\end{thebibliography}

\end{document}